\documentclass[]{sn-jnl}

\usepackage{amsmath}
\usepackage{empheq}
\usepackage{mathrsfs}
\usepackage{geometry}                
\usepackage{graphicx}
\usepackage{amssymb}
\usepackage{epstopdf}
\usepackage{graphicx}%
\usepackage{multirow}%
\usepackage{amsmath,amssymb,amsfonts}%
\usepackage{amsthm}%
\usepackage{mathrsfs}%
\usepackage[title]{appendix}%
\usepackage{xcolor}%
\usepackage{textcomp}%
\usepackage{manyfoot}%
\usepackage{booktabs}%
\usepackage{algorithm}%
\usepackage{algorithmicx}%
\usepackage{algpseudocode}%
\usepackage{listings}%
\usepackage{natbib}%
\newcommand{\G}{\mathcal{G}}
\newcommand{\M}{M_{\odot}}
\newcommand{\C}{\mathcal{C}}
\newcommand{\J}{\mathcal{J}}
\newcommand{\Ham}{\mathcal{H}}
\newcommand{\Mioc}{M_{\rm{IOC}}}
\newcommand{\as}{\tilde{a}}
\newcommand{\bs}{\tilde{b}}

\begin{document}

\title[Self-Gravitational Dynamics Within the Inner Oort Cloud]{Self-Gravitational Dynamics Within the Inner Oort Cloud}

\author*[1]{\fnm{Konstantin} \sur{Batygin}}\email{kbatygin@caltech.edu}

\author[2]{\fnm{David} \sur{Nesvorn\'y}}\email{davidn@boulder.swri.edu}

\affil*[1]{\orgdiv{Division of Geological and Planetary Sciences}, \orgname{California Institute of Technology}, \orgaddress{\street{1200 E. California Blvd.}, \city{Pasadena}, \postcode{91125}, \state{CA}, \country{USA}}}

\affil[2]{\orgdiv{Department of Space Studies}, \orgname{Southwest Research Institute}, \orgaddress{\street{1050 Walnut St., Suite 300}, \city{Boulder}, \postcode{80302}, \state{CO}, \country{USA}}}


\abstract{The formation of the Inner Oort Cloud (IOC) — a vast halo of icy bodies residing far beyond Neptune's orbit — is an expected outcome of the solar system's primordial evolution within a stellar cluster. Recent models have shown that the process of early planetesimal capture within the trans-Neptunian region may have been sufficiently high for the cumulative mass of the Cloud to approach several Earth masses. In light of this, here we examine the dynamical evolution of the IOC, driven by its own self-gravity. We show that the collective gravitational potential of the IOC is adequately approximated by the Miyamoto-Nagai model, and use a semi-analytic framework to demonstrate that the resulting secular oscillations are akin to the von Zeipel-Lidov-Kozai resonance. We verify our results with direct $N$-body calculations, and examine the effects of IOC self-gravity on the long-term behavior of the solar system's minor bodies using a detailed simulation. Cumulatively, we find that while the modulation of perihelion distances and inclinations can occur within an observationally relevant range, the associated timescales vastly surpass the age of the sun, indicating that the influence of IOC self-gravity on the architecture of the solar system is negligible. }


\maketitle

\section{Introduction}

The prevailing consensus holds that our solar system originated within a stellar cluster -- a conclusion bolstered by multiple lines of evidence that extend beyond the statistical observation that most stars do not form in isolation \citep{Adams2010}. A key piece of supporting evidence for this notion is the prevalence of decay products from short-lived radionuclides (notably $^{26}$Al), within primitive bodies of the solar system. The widespread presence of such elements is commonly viewed as a strong indicator that the proto-solar cloud emerged in a relatively densely populated stellar association \citep{2001Icar..150..151A,2023A&A...670A.105A}.

The presence of a stellar cluster around the nascent solar system had a number of important implications, influencing not just the isotopic makeup of solar system bodies but also their dynamics. In fact, the formation of the Inner Oort Cloud (IOC) — a population of long-period minor bodies that do not strongly interact with the giant planets or the Galaxy — necessitates extrinsic perturbations \citep{1981AJ.....86.1730H,Fernandez1997}. Particularly, the assembly of the IOC is envisioned to have taken place through the following sequence of events. First, as giant planets accreted and migrated within the proto-solar nebula, they scattered primordial planetesimals onto highly eccentric orbits. Over time, these orbits chaotically diffused outward, where the combined effects of stellar flybys and the cluster's tidal forces began to alter their trajectories. For a subset of these objects, external perturbation led to orbital \textit{circularization}, dynamically detaching them from the giant planets. Eventually, as the sun exited the cluster, these orbits became fossilized in the trans-Neptunian region of the solar system, generating a population of objects with heliocentric distances that span hundreds to thousands of AU \citep{Brasser2006,Brasser2012,Kaib2008}.

In light of the expected existence of a vast population of IOC objects, several authors have explored the potential role of self-gravitational effects in shaping the large-scale architecture of the trans-Neptunian region. For instance, \citet{2016MNRAS.457L..89M, 2023ApJ...948L...1Z} have examined the emergence of the so-called inclination instability, which occurs due to the self-interaction of a multitude of highly eccentric orbits initially confined to a common plane. Their findings suggest that non-trivial dynamical behavior could arise on a timescale far exceeding the orbital period, provided the combined mass of the orbiting bodies exceeds $\sim10$ Earth masses. However, the recent study of \citet{2023MNRAS.523.6103D} showed that the inclusion of self-consistent modeling of the giant planets -- particularly Neptune’s scattering effects -- fully suppresses the inclination instability.

In a separate theoretical approach, \citet{2019AJ....157...59S} investigated the collective effects of a massive, lopsided trans-Neptunian disk of planetesimals extending to hundreds of AU. The authors proposed that such a disk could account for the observed clustering of longitudes of perihelion among long-period, dynamically detached TNOs \citep{2016AJ....151...22B}. Yet, the physical mechanisms responsible for the formation and maintenance of such an asymmetric structure in the outer solar system remain elusive.

Recently, \citet{bib1} presented a detailed study of the IOC's formation, which incorporates sophisticated modeling of the early migration of the giant planets, along with a self-consistent treatment of the cluster's effects. Through subsequent $N-$body simulations, they showed that forming the solar system in a dense stellar environment produces perihelion and semimajor axis distributions among long-period trans-Neptunian objects that closely mirrors the observational data. Intriguingly, these simulations also indicate that the total mass captured in the IOC can exceed the mass of the Earth by as much as a factor of three, a revelation that underscores the efficiency with which cluster-driven dynamics could have shaped the outer reaches of the solar system.

In a distinct effort, \citet{2019A&A...629A..95S} considered the secular evolution of the IOC under the combined perturbations from the giant planets and the Galactic tide. Remarkably, they demonstrated that much of the parameter space associated with the IOC is perforated by chaos, insinuating that the long-term evolution of the IOC is considerably less inert than previously thought (although the rate of chaotic diffusion is exceptionally slow).

Inspired by the combined works of \citet{2019A&A...629A..95S} and \citet{bib1}, here we explore previously overlooked dynamics within the IOC that are driven by its own self-gravity. To this end, in Section 2, we introduce the Miyamoto-Nagai potential-density pair as a simple model for the collective gravity of the IOC. Then, in Section 3, we demonstrate from semi-analytic grounds that the emergent secular interactions are akin to von Zeipel-Lidov-Kozai cycles. Subsequently, we show that while self-gravity can, in principle, drive dynamically detached orbits to join the Scattered Disk, the timescale to achieve this greatly exceeds the lifetime of the solar system, unless the mass of the IOC is assumed to be unjustifiably large.

\begin{figure*}[t]
\centering
\includegraphics[width=\textwidth]{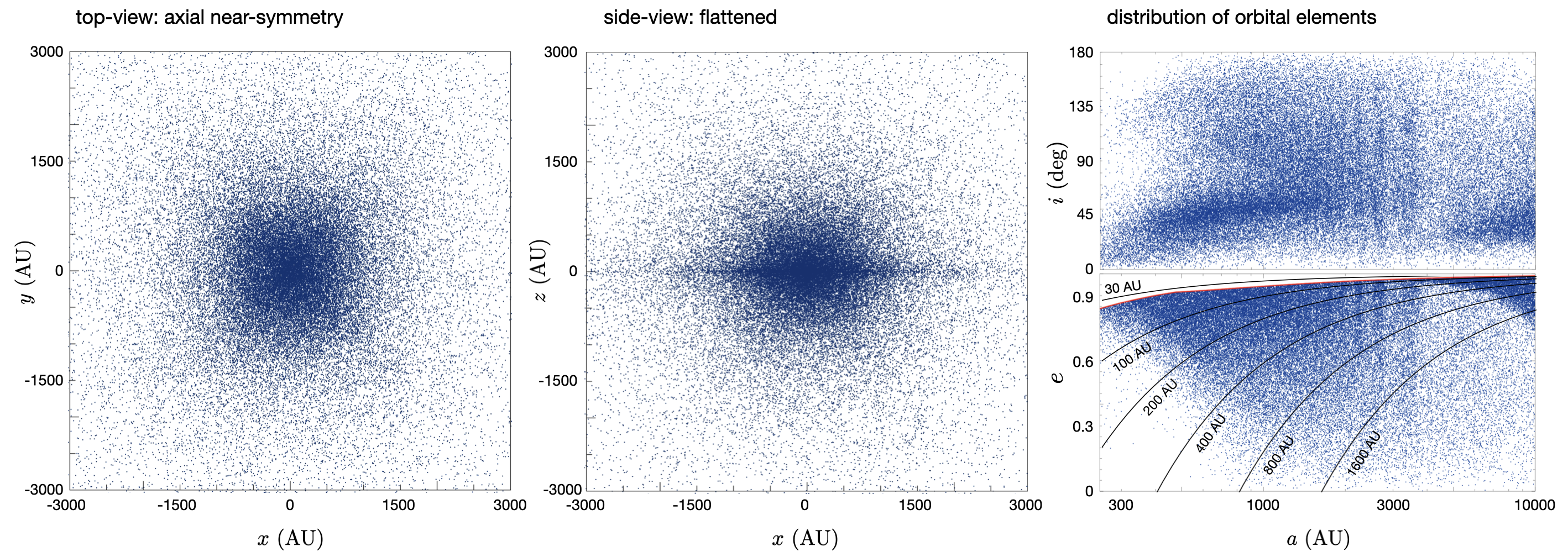}
\caption{Visualization of the particle distribution within the IOC as derived from the $t=300\,$Myr snapshot of the \citet{bib1} \texttt{cluster2} simulation. Only objects with semi-major axes in excess of 250 AU and perihelia greater than the scattering threshold given by equation (\ref{qcrit}) are shown. The left panel displays a top-down view (x-y projection) of the particle distribution, highlighting the approximate axisymmetry of the generated cloud. The middle panel presents the side view (x-z projection) of the same distribution, effectively illustrating the cloud's dual composition: a central flattened disk is embedded within a more extended, quasi-spherical halo. The right panel depicts the distribution of orbital elements of the IOC: the inclination (top) and eccentricity (bottom) are shown as functions of the semi-major axis, with the bounding (critical) perihelion distance (equation \ref{qcrit}) marked with a red curve. Additionally, the semi-major axis -- eccentricity panel depicts contours of constant perihelion distance, with the values of $q$ labeled accordingly.}
\label{f1}
\end{figure*}

\section{Miyamoto-Nagai Model of the Inner Oort Cloud}

As a first step in our analysis, let us consider the structure of the IOC. To construct the physical model, we derive the initial conditions from the \texttt{cluster2} simulation presented in \citet{bib1}, adopting the distribution of IOC objects at the 300 Myr mark. This simulation provides a nuanced model of the solar system's primordial evolution, incorporating detailed and well-tested prescriptions of giant planet migration and the dynamical influence of the Sun's birth cluster. Specifically, the \texttt{cluster2} simulation models stellar encounters self-consistently, assuming a Plummer model for the Sun’s birth cluster characterized by a mass of 1200$M_{\odot}$ and a Plummer radius of 0.35 pc. This environment approximates conditions akin to the Orion Nebular Cluster (see e.g, \citealt{Bat20}), and the product of the local stellar density and the sun's residence time within the cluster -- taken to be $\eta\,\tau\sim10^4$Myr pc$^{-3}$ -- is tuned to reproduce the semi-major axis distribution of high-perihelion scattered disk objects.

The choice of the 300 Myr snapshot is informed by the timeline of early solar system evolution and cluster dynamics. Despite being early in the the solar system’s lifetime, by this epoch, the giant planets' migration — modeled in \citet{bib1} with a characteristic migration timescale of $\tau_{\rm{mig}} = 10$ Myr has largely concluded. Additionally, this point marks a phase where the solar system's birth cluster had already dissipated, and the formation of the IOC is complete.

Objects on unstable\footnote{See e.g., \citet{2019PhR...805....1B, 2024ApJ...962L..33H} for long-term integrations of unstable large-semi-major axis TNOs.} (i.e., Neptune-crossing and scattering) orbits will diffuse out of the solar system on a sub-Gyr timescale, and will therefore not contribute to the cumulative gravitational potential of the (long-term stable) “inert” inner Oort Cloud. Indeed, in the limit where $q \rightarrow a_{\rm{N}}$, the semi-major axis diffusion approaches $\mathcal{D}_a\sim 8/(5\pi)\,(m_{\rm{N}}/\M)\,\sqrt{\G\,\M\,a_{\rm{N}}}$ \citep{Bat21,2024MNRAS.527.3054H}. In an effort to filter these particles out of our analysis and isolate the relevant population within the broader trans-Neptunian region, we exclude particles within the active Neptune-scattering zone, as well as those with (comparatively) short periods. This is achieved by removing objects with semi-major axis smaller than $250\,$AU and perihelion distance below the scattering threshold \citep{Bat21}:
\begin{align}
q_{\rm{crit}} = \mathrm{sup} \Bigg[q_0, a_{\rm{N}} \sqrt{\log \left(\frac{24^2}{5} \frac{m_{\rm{N}}}{\M} \left(\frac{a}{a_{\rm{N}} }\right)^{5/2}\right)} \Bigg],
\label{qcrit}
\end{align}
where quantities with the subscript ``N" refer to Neptune, and $q_0=40\,$AU. It is important to note that this expression for the critical perihelion distance is derived under the assumption of planar motion ($i\rightarrow0$) and small semi-major axis ratio ($a_{\rm{N}}/a\ll1$) (\citealt{Bat21}; see also \citealt{2024MNRAS.527.3054H}), and hence, our approach may be somewhat restrictive. That is, chaotic diffusion of orbits is reduced at high inclinations, suggesting that our cut-off might be more stringent than necessary, given that it is derived in the planar approximation. However, for the purposes of our initial exploration into the IOC's dynamics, this simplification provides a suitable starting point.

\begin{figure*}[t]
\centering
\includegraphics[width=\textwidth]{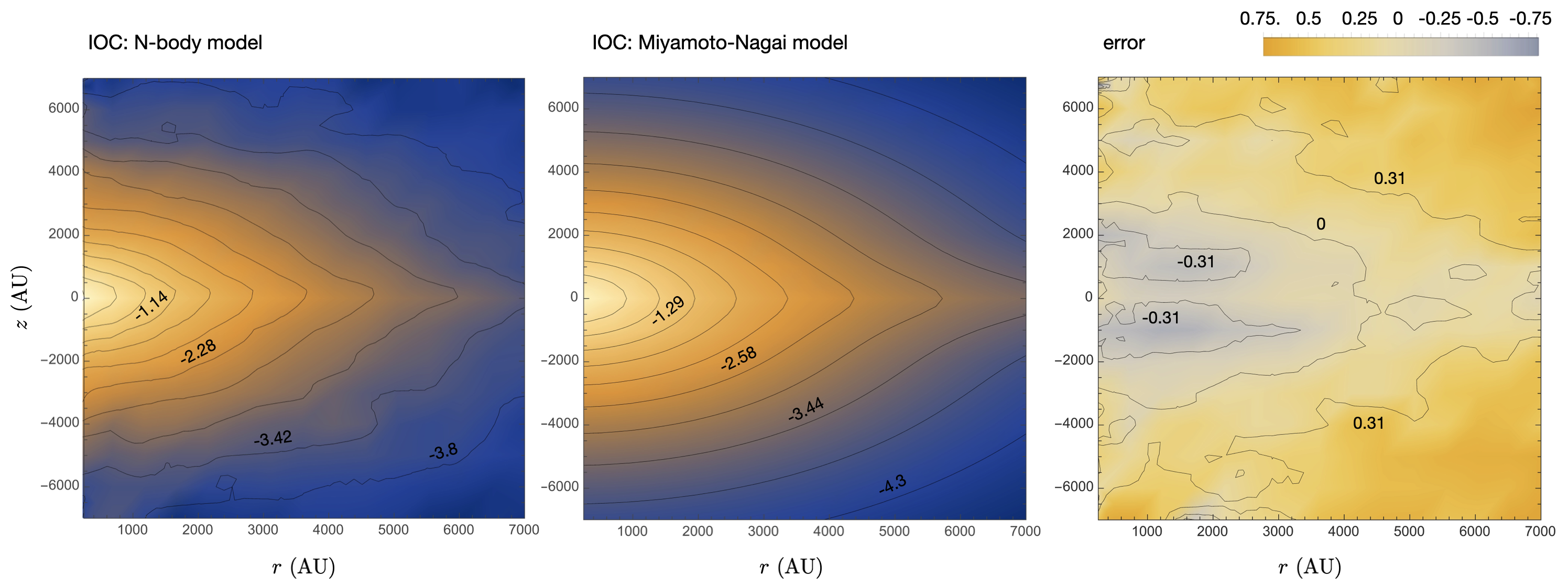}
\caption{Density profile of the IOC. The left panel showcases the azimuthally-averaged particle density, $\rho_{\rm{num}}$ derived from detailed $N-$body simulation of \citealt{bib1} (see also Fig. 1). The middle panel shows the Miyamoto-Nagai density profile, $\rho$ (given by equation \ref{psi}), with scaling parameters $\as=200\,$AU and $\bs/\as=5$. In both panels, the density is normalized to unity at $r=250$ AU and $z=0$, and a sequence of contours is labeled, depicting $\log_{10}(\rho)$. The right panel depicts the error associated with the analytic fit, $\epsilon = (\rho-\rho_{\rm{num}})/\rho_{\rm{num}}$. As can be seen, the error within the relevant regions is limited to tens of percent, which is satisfactory for the purposes of our study.}
\label{f2}
\end{figure*}

Examination of the particle distribution in physical space (as shown in Fig 1) reveals that the particle distribution within the IOC is, to a good approximation, axisymmetric\footnote{To ascertain the degree of axial symmetry, we computed the distribution of azimuthal angles of particles across cylinders with a radial thickness of 100 AU at radial intervals of 100 AU, and applied the Kolmogorov-Smirnov test for uniformity. This calculation yielded an average $p$-value of $\langle p\rangle=0.45$, indicating that axial symmetry is statistically consistent with the simulation data.}. However, the distribution is markedly not spherical: when viewed from the side, there is a discernible disk-like structure embedded within a more diffuse halo. To model this distribution, we adopt the Miyamoto-Nagai potential-density pair \citep{1975PASJ...27..533M}. 

The Miyamoto-Nagai potential is a versatile model that allows for the simulation of a broad range of astrophysical systems, with adjustable thickness and radial distribution of mass. It is defined by two crucial parameters: the radial scale length ($\as$) and the scale height ($\bs$). These parameters govern the shape and density distribution of the particle population. As $\bs$ approaches zero, the model converges to the Kuzmin potential, representing a razor-thin disk. Conversely, when $\as$ approaches zero, the model reduces to a Plummer sphere. The ratio $\bs/\as$ is thus a key factor in determining the disk's flatness: a high $\bs/\as$ ratio results in a mass distribution that is nearly spherical, while a low $\bs/\as$ ratio leads to a significantly flattened distribution. This flexibility allows the Miyamoto-Nagai potential to adeptly model various astrophysical structures by adjusting $\as$ and $\bs$ accordingly.


Written in terms of cylindrical coordinates, the potential $\Psi$ and the corresponding density have the functional form:
\begin{align}
\label{psi}
\Psi &= -\frac{\G\,\Mioc }{\sqrt{\big(\as+\sqrt{\bs^2+z^2}\big)^2+r^2}}, \\ \nonumber
\rho &= \frac{\nabla^2\,\Psi}{4\,\pi\,\G} = \frac{\bs^2 \Mioc\, \big( \big(\as+\sqrt{\bs^2+z^2}\big)^2 \big(\as+3 \sqrt{b^2+z^2}\big)+ \as \,r^2\big)}{4\,\pi\,\big(\bs^2+z^2\big)^{3/2}  \big(\big(\as+\sqrt{\bs^2+z^2}\big)^2+r^2\big)^{5/2}},
\end{align}
where $\Mioc$ is the total mass of the cloud. 

To determine the appropriate parameter values, we began by fitting the analytic model to particle distribution shown in Fig 1, which yielded $\as=155$ AU and $\bs/\as=7$. We subsequently fine-tuned these parameters to $\as = 200$ AU and $\bs/\as=5$. This adjustment slightly reduced the discrepancy between the analytic model and the numerical results within the disk region, albeit at the cost of accuracy in the much more sparsely populated regions at very large heliocentric distances. Figure 2 depicts a comparison between the azimuthally-averaged particle distribution obtained from the detailed $N-$body simulation and that given by equation \ref{psi}.


\section{Secular Dynamics Within the IOC}

With the potential of the IOC now defined, we are in a position to write down the model Hamiltonian that will govern the dynamics within this system. The chosen potential, characterized by its smoothness and axisymmetry, implies that the IOC will predominantly drive secular evolution. This class of dynamics corresponds to long-term orbital changes, smoothing over short-term perturbations. It is important to acknowledge, however, that this phase-averaged description, only provides an adequate approximation for bodies outside the scattered disk, and more generally, for objects not trapped in high-order MMRs with Neptune \citep{2002mcma.book.....M}.

Accounting for secular perturbations from Jupiter, Saturn, Uranus, and Neptune to quadrupolar order, we have:
\begin{align}
\label{Ham}
\Ham = \frac{\G\,\C\,\big(3\cos^2(i) - 1 \big)}{8\,a^3\,\big(1-e^2 \big)^{3/2}} + \frac{1}{2\,\pi}\oint\Psi\,d\mathcal{M},
\end{align}
where $\mathcal{M}$ is the mean anomaly of the IOC particle, and $\C=\sum_i m_i\,a_i^2$ is the sum of the moments of inertia of the planetary orbits. Although the integral in the above expression cannot be evaluated analytically, general properties of $\Ham$ can nevertheless be deduced from symmetry. First, $\Ham$ is, by construction, independent of the orbital phase, meaning that the semi-major axis, $a$, is rendered a constant of motion. Similarly, axial symmetry implies that $\Ham$ is independent of the longitude of ascending node, $\Omega$, meaning that the conjugate action, $\J$, -- which corresponds to the $\hat{z}$-component of the specific angular momentum vector, is conserved. In other words, $\Ham = f(a,\J,e,\omega)$ depends only on a single angle -- the argument of perihelion, $\omega$. Moreover, because the difference between ascending and descending nodes disappears upon phase-averaging, the dependence of $\Ham$ on $\omega$ is bound to be $\pi-$periodic.

These properties place $\Ham$ into the broad category of von Zeipel-Lidov-Kozai type Hamiltonians, i.e., systems with only a single degree of freedom -- related to the $(e,\omega)$ variable pair -- parameterized by $a$ and $\J$ \citep{1910AN....183..345V,2023MNRAS.522..937T}. Thus, upon an appropriate choice of the constants of motion, the phase-space portrait of an IOC particle can be fully quantified by projecting level curves of $\Ham$ onto the $(\omega,q)$ plane.

\begin{figure*}[t]
\centering
\includegraphics[width=0.85\textwidth]{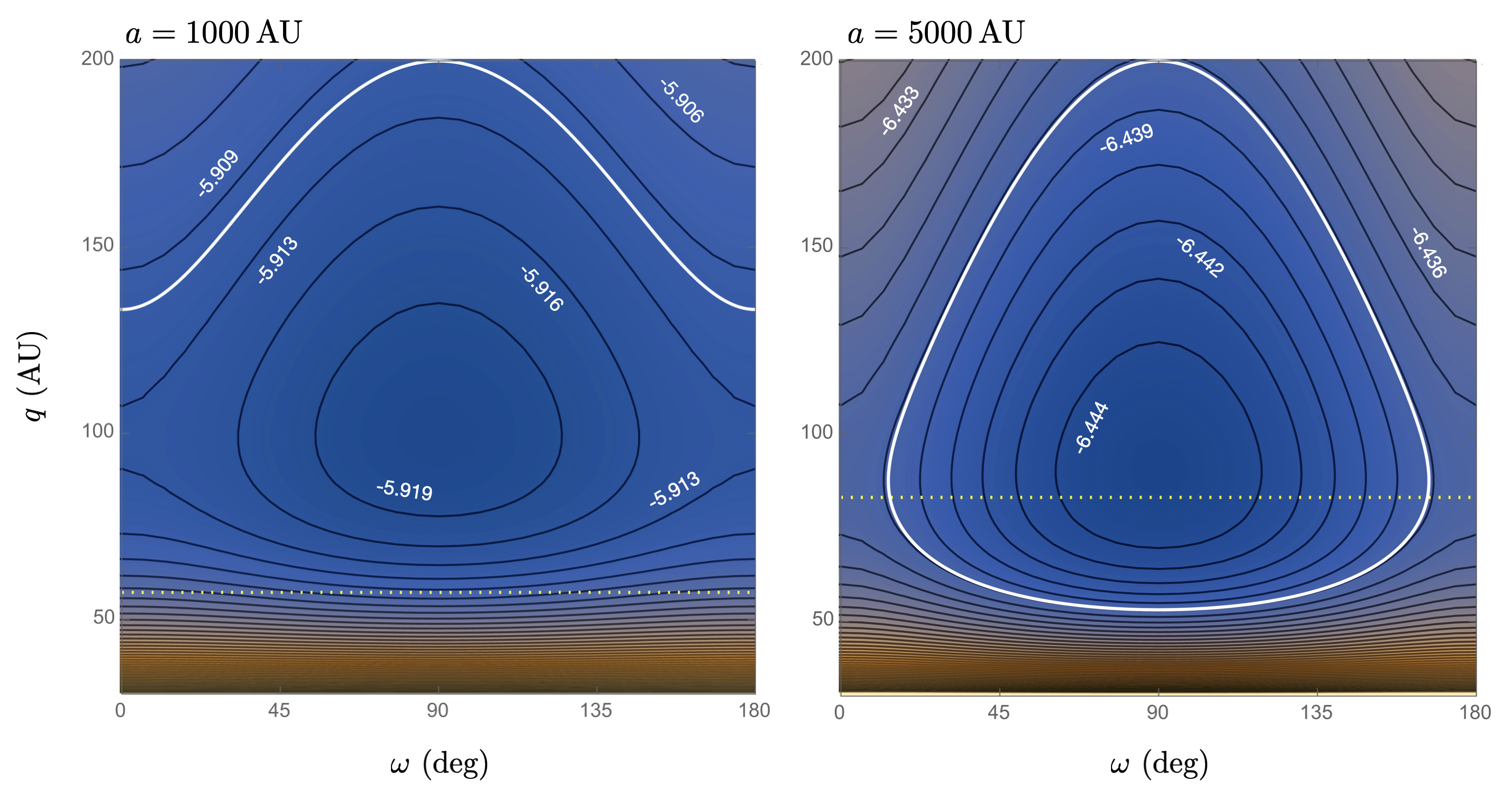}
\caption{Level curves of Hamiltonian (3) projected onto the $(\omega,q)$ plane. These contours depict the secular dynamics within the IOC for particles with semi-major axes of $a=1000\,$AU (left panel) and $a=5000\,$AU (right panel), for a total IOC mass of $\Mioc=3M_{\oplus}$. Both panels illustrate the characteristic structure of von Zeipel-Lidov-Kozai (vZLK) resonance. A series of contours are labeled according to $\log_{10}(-\mathcal{H})$, evaluated in units of $M_{\odot}$, AU, and yr. The dotted yellow line corresponds to the Neptune-scattering threshold in the coplanar limit, and the white curves depict output from an $N-$body simulation, for equivalent values of $\J$.}
\label{f3}
\end{figure*}


Though the IOC as a whole covers a broad range of parameters, here, our selection of constants of motion is motivated by the observational census of known long-period TNOs, which primarily consists of high-eccentricity objects situated near the plane of the solar system. Considering the perturbative nature of the model, the maximum allowable eccentricity corresponds to a perihelion distance equal to Neptune's semi-major axis, beyond which orbits would cross and fundamentally disrupt the model's accuracy. Thus, for definitiveness, we define the action variable to correspond to an orbit with zero inclination and $q = a_{\rm{N}}$:
\begin{align}
\label{Ham}
\J = \sqrt{1-e^2}\,\cos(i) = \sqrt{\frac{a_{\rm{N}}\,(2\,a-a_{\rm{N}})}{a^2}} = \mathrm{const.}
\end{align}
We reiterate, however, that even for orbits with larger perihelion distances, the secular Hamiltonian can yield an inaccurate representation of the dynamics, if the IOC particle becomes embedded within a dense network of mean motion resonances that facilitate Neptune scattering \citep{Bat21}. Despite these limitations, this semi-analytic approach is still effective in exploring whether orbital evolution driven by IOC’s self-gravity, can elevate the eccentricities of initially high-perihelion objects to the scattering threshold, thereby cycling them between the IOC and scattered disk.

\begin{figure*}[t]
\centering
\includegraphics[width=0.85\textwidth]{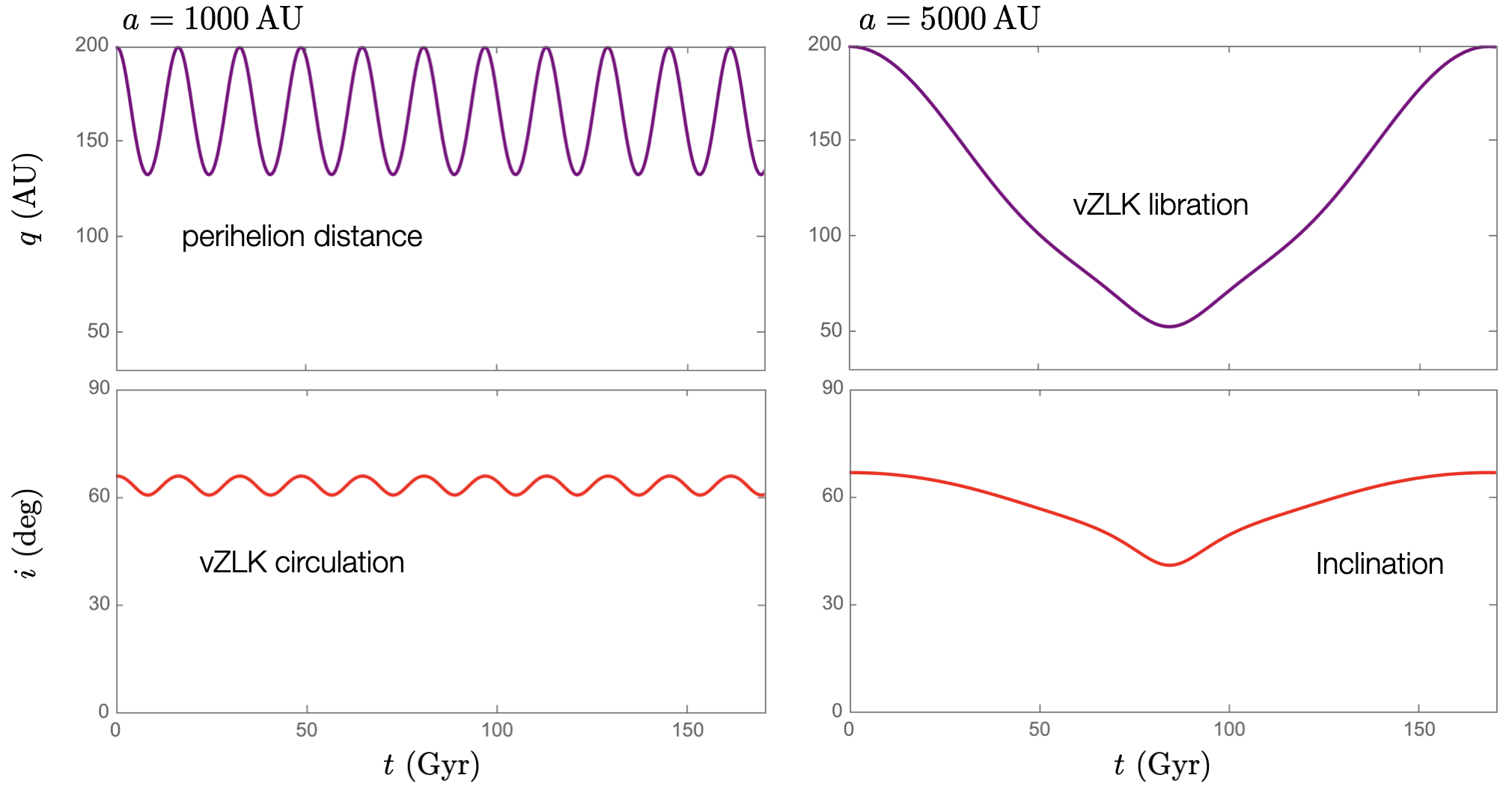}
\caption{Time-series of perihelion distance and inclination from idealized N-body simulations of IOC particles at $a=1000\,$AU (left panel) and $a=5000\,$AU (right panel) for $\Mioc=1M_{\oplus}$. Driven by IOC's self-gravity, the particles execute vZLK dynamics, where perihelion distance and inclination oscillate in concert, preserving the vertical component of the specific angular momentum vector. These numerically computed orbits are projected onto the $(\omega,q)$ in Figure 3, providing a direct comparison between the simulation results and the analytical model. While the agreement between the two is satisfactory, these time-series also underscore the exceedingly long timescales over which IOC-driven secular dynamics operate. As depicted, the orbital evolution unfolds over durations that far exceed the age of the solar system.}
\label{f4}
\end{figure*}

Figure 3 shows the contours of $\Ham$ on the $(q,\omega)$ plane for IOC particles at $a=1000$ and $5000\,$AU, where we have set $\Mioc=3M_{\oplus}$, in agreement with the simulation results of \citet{bib1}. In both panels, the familiar structure of the vZLK resonance can be seen, and intriguingly, secular IOC-driven dynamics appears to modulate the perihelia over an observationally relevant (i.e., $q\sim100\,$AU) range. Moreover, the reduction of the phase-space portrait towards trivial circulation for $q$ approaching $30$ AU can be understood as emanating from the growing contribution of the first term in Hamiltonian (3). More specifically, our analysis demonstrates that the contribution of the first term, representing precession driven by the giant planets, to the overall secular dynamics of the IOC, significantly varies with perihelion distance. At $a=1000$ AU, this term accounts for less than 2\% of the Hamiltonian's value at $q=75$ AU, increasing to 7\% and 25\% at $q=50$ and $30$ AU, respectively. Meanwhile, at $a=5000$ AU, its influence is relatively subdued, constituting approximately 10\% at $q=30$ AU and diminishing to 2\% at $q=50$ AU, thereby allowing the vZLK cycle to penetrate to smaller perihelion distances.

To further quantify these self-gravitational dynamics, we supplement our analytic calculations with direct numerical orbit propagation of test particles. To carry out the integrations, we used the \texttt{mercury6} gravitational software package \citep{1999MNRAS.304..793C}, utilizing the MVS algorithm \citep{1991AJ....102.1528W}. To ensure consistency with our analytical model, we include the accelerations due to the IOC via $\mathbf{a}=-\nabla\,\Psi$ and model the quadrupolar fields of the giant planets as an oblateness of the Sun, represented by $J_2 = \C/(2\,\M\,\mathcal{R}^2)$ moment, setting the central body's radius (which also acts as an absorbing boundary condition to $\mathcal{R}=a_{\rm{N}}$). The initial conditions for the IOC objects, treated as test particles, are set with a perihelion of 200 AU, an argument of perihelion of $\omega=90\deg$, and an initial inclination determined by the constant of motion, $\J$.

The time-series of the perihelion distance and inclination are shown on Figure 4. The numerically computed evolution of the orbits is also projected onto the ($\omega,e$) diagrams shown in Figure 3. Overall, these simulations show excellent agreement with the semi-analytic model. However, the time-series also highlight the staggeringly long timescales associated with the evolution driven by the self-gravity of the IOC. In fact for the adopted parameters, evolution unfolds on a timescale comparable to the age of the universe or even longer.

Given the extraordinarily long integration period, we clarify that the intent behind these simulations is not to present a realistic portrayal of the Solar System's future evolution. Instead, this calculation merely serves as an illustrative approach to fully capture the dynamical cycle of particles evolving under the influence of the inner Oort Cloud’s self-gravitational potential. Indeed, owing to external perturbations such as stellar flybys, the inner Oort cloud is expected to be depleted on a timescale much shorter than that depicted in Figure 4. As importantly, the dynamical lifetime of the outer solar system itself is limited to mere tens of Gyr \citep{2020AJ....160..232Z}. Accordingly, these simulations should be viewed as nothing more than a theoretical exposition of the isolated dynamics facilitated by Hamiltonian (3).

\section{N-body simulations}

\begin{figure}[h!]
\centering
\includegraphics[width=0.6\columnwidth]{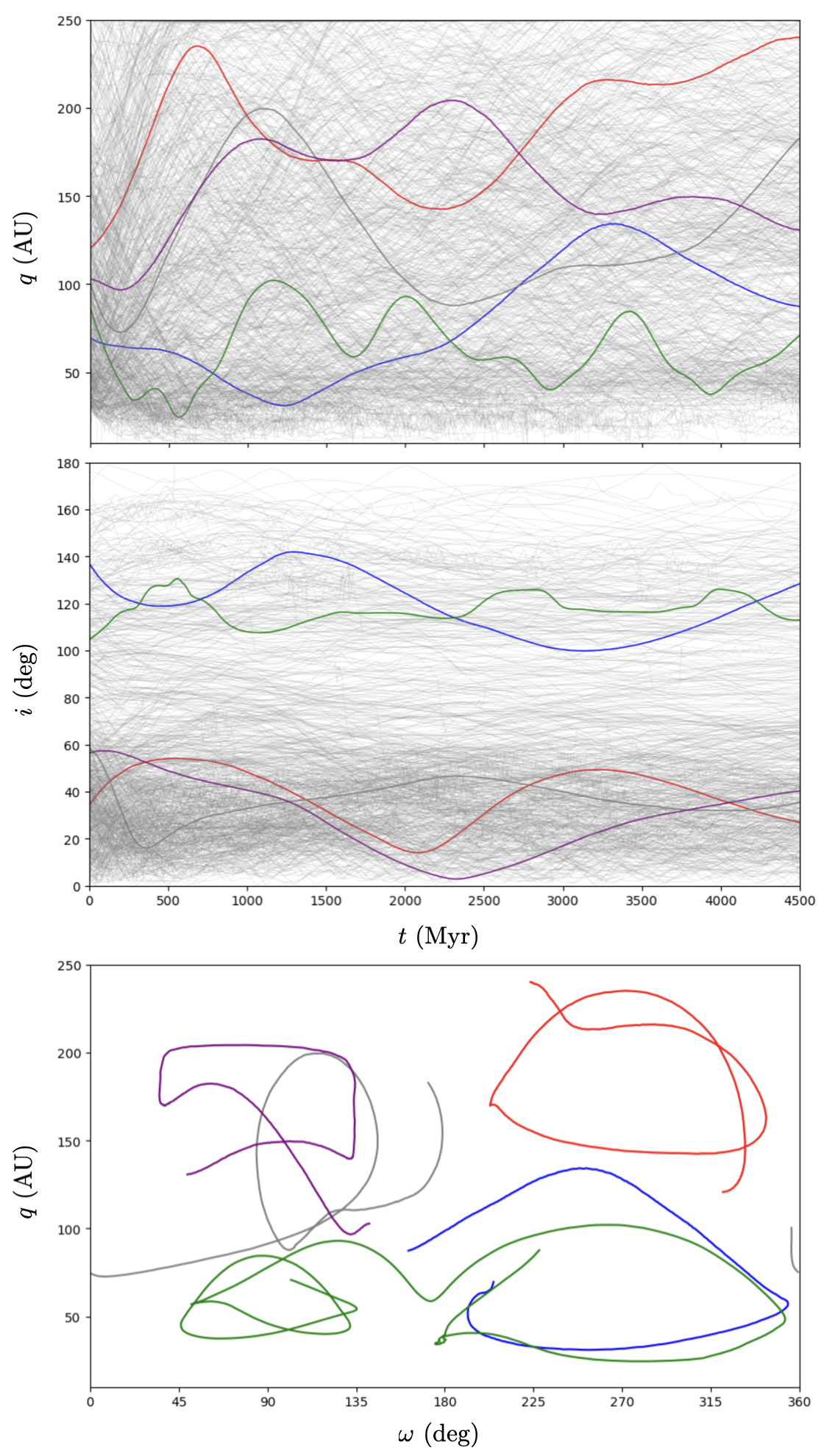}
\caption{Time-series of perihelion distance (top panel) and inclination (middle panel) of IOC objects, derived from the $N=10^{4}$ \texttt{GENGA} simulation. A subset of trajectories that are entrained in vZLK cycles are highlighted, emphasizing long-term evolution of such particles. Notably, particles entrained in cycles of $\omega-$libration are rare (comprising $\sim2\%$ of all objects), but their existence draws a parallel between direct $N$-body simulations and the semi-analytic model based upon the Miyamoto-Nakai potential. The time-series of these objects are further illustrated in the $q-\omega$ diagram (bottom panel), confirming the libration of the argument of perihelion and showcasing the vZLK behavior expected from the semi-analytic model.}
\label{f5}
\end{figure}

In an effort to explore the self-gravitational dynamics of the IOC further, we conducted a pair of additional N-body simulations. First, as a test of the secular framework outlined in the previous section, we used the \texttt{GENGA} GPU-accelerated code \citep{2022ApJ...932..124G} to directly model the gravitational interactions, without assuming the Miyamoto-Nagai potential. The principal goal of this calculation was to assess if the vZLK-type behavior insinuated by the semi-analytic model can be recovered within the context of a direct (``brute force") calculation.

The initial conditions for our simulation were directly derived from the distributions shown in the right panel of Figure 1. To expedite the dynamical evolution, however, we implemented two significant adjustments: firstly, we augmented the cloud's mass and secondly, we consolidated its distribution into a denser orbital configuration. In particular, we excluded all objects with semi-major axes greater than 1000 AU and perihelia below $q<35\,$AU, while elevating the total mass of the remaining distribution to 10 Earth masses. This reduction in the range of semi-major axes (and orbital periods), coupled with the substantial enhancement in total mass, significantly intensified the gravitational interactions within the system. The perturbations of Jupiter, Saturn, and Uranus were encapsulated within an effective $J_2$ oblateness coefficient as before, while Neptune was explicitly included on its current orbit. The time-step for the simulation was set at 5 years.

The simulation spanned a total integration time of 4.5 billion years, mirroring the age of the solar system. The results, presented in Figure 5, elucidate the long-term evolution of the particles. In particular, the top two panels of Figure 5 display the time-series of perihelion distances and inclinations, respectively. The bottom panel of Figure 5 shows the perihelion distance as a function of the argument of perihelion for a subset of the simulated particles. This diagram serves as a direct comparison to the semi-analytic model presented earlier in Figure 3, highlighting the libration zones characteristic of vZLK resonance. Though the depicted evolution is notably stochastic (owing in part to the relatively coarse representation of the IOC), the behavior of these particles aligns with the anticipated vZLK dynamics driven by the IOC's self-gravity.

\begin{figure}
\centering
\includegraphics[width=0.75\columnwidth]{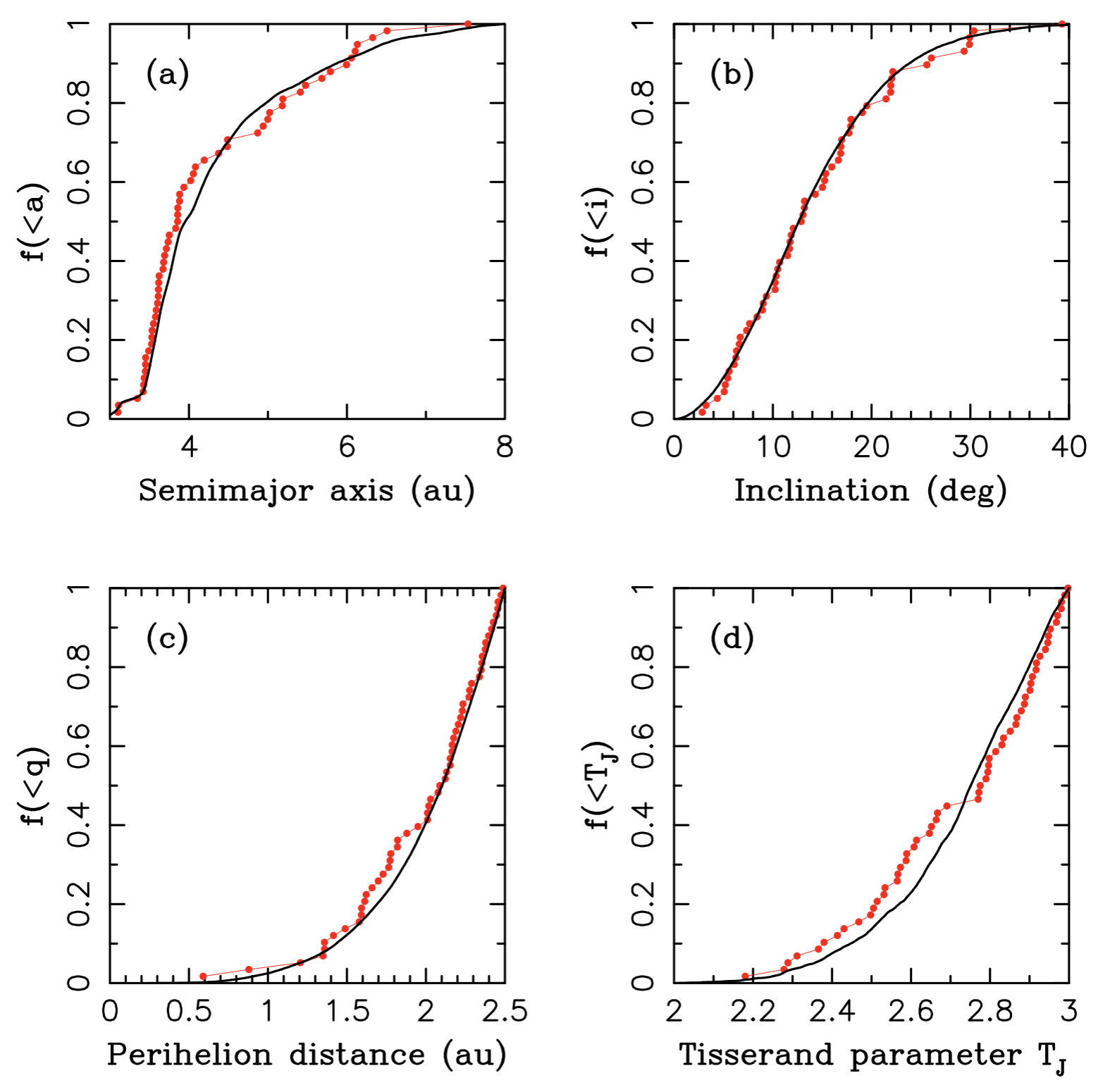}
\caption{Comparison of observed and simulated cumulative distributions for orbital elements and the Tisserand parameter of Jupiter Family Comets (JFCs). Panels (a) to (d) show the cumulative fraction $f$ of JFCs for semi-major axis, inclination, perihelion distance, and Tisserand parameter with respect to Jupiter ($T_{\rm{J}}$), respectively. The red dots indicate observations (from the minor planet center), while the solid black line represents results from our simulation incorporating self-gravity with a 3 Earth mass IOC. This figure closely resembles the self-gravity-free simulation presented in \citet{2017ApJ...845...27N} underscoring the negligible influence of self-gravity on the orbital evolution of JFCs and the solar system as a whole.}
\label{f6}
\end{figure}


Having not observed any obvious contradictions between the $\texttt{GENGA}$ simulations and the dynamical behavior expected within the context of the orbit-averaged Miyamoto-Nagai framework, we revisited the calculations of \citet{bib1}, incorporating the potential (\ref{psi}) as an auxiliary effect, and retaining the IOC mass at $3 M_{\odot}$. Broadly speaking, this simulation revealed no significant difference in any of the readily-observable trans-Neptunian populations of minor bodies within the solar system, we additionally interrogated the cometary population for any hints of difference between runs with and without self-gravity. While the perturbations associated with IOC self-gravity are confined to large heliocentric distances, short period comets within the solar system constitute an observationally well-quantified aggregate of minor bodies that are derived from the trans-Neptunian region (namely, they originate at large orbital periods but scatter inwards). As a result, perturbations to the outer solar system naturally propagate to the cometary populations, and their distribution has previously been shown to be affected by perturbations from dwarf planets \citep{Munoz19} as well as the hypothetical Planet 9 \citep{2017ApJ...845...27N}.


We examined the generation of Jupiter Family Comets (JFCs) in the last billion years as a  proxy for the effects of self-gravity. The results, presented in Figure 6, allow for direct comparison with the findings of \citet{2017ApJ...845...27N}, which did not account for the cluster potential. As expected, inclusion of IOC self-gravity did not meaningfully alter the production of JFCs compared to the published results of \citet{2017ApJ...845...27N}. The fact that this simulation does not violate the JFC constraint is consistent with the aforementioned notion that the timescales for self-gravitational dynamics within the IOC are too long to substantially impact the solar system's evolution. Indeed, the muted effect of self-gravity aligns with theoretical expectations outlined in the preceding section, and underscores its ultimately negligible role in shaping the dynamical architecture of the trans-Neptunian region.


\section{Discusssion}



In our study, we have considered a self-consistent model for the formation of the IOC to analyze the secular dynamics of objects within the IOC, under the influence of their collective gravitational potential. By aligning a Miyamoto-Nagai potential-density pair with the structure observed in simulation outputs, we quantified the von Zeipel-Lidov-Kozai-type dynamics among dynamically detached orbits. Our findings even suggest that the range of perihelion modulation facilitated by the vZLK cycles can reach perihelion distances of $q\lesssim100$ AU, where the ``detached" population of scattered disk objects is observed. However, our simulations also established that the timescale for these self-gravitational dynamics to manifest significantly exceeds not only the solar system's age but also the expected dynamical lifetime of the giant planets (see \citealt{2020AJ....160..232Z}).

Although our analysis is predicated on a specific scenario for the IOC's formation, and alternative model parameters could theoretically yield quantitatively different outcomes, this is unlikely because the particular simulation results of \citet{bib1} adopted in this study already invoke cluster parameters and solar residence time that are close to the upper bound of the constraints set by the dynamical structure of the cold classical belt \citep{Bat20}. This means that more massive models of the IOC are likely to be incompatible with other solar system constraints. Our results thus suggest that any realistic parameter range would likely lead to similar conclusions. Consequently, our research joins the ranks of other unsuccessful attempts to attribute the dynamical architecture of the outer solar system to self-gravitating effects. 

\bmhead{Acknowledgements}

We are thankful to Fred Adams, Alessandro Morbidelli, Cristian Beauge, and Mike Brown for insightful discussions. We thank the anonymous referees for providing careful and insightful reviews of the manuscript. K.B. is grateful to Caltech, the David and Lucile Packard Foundation, and the National Science Foundation (grant number: AST 2109276) for their generous support.

\bmhead{Conflict of Interest}
The authors declare that they have no competing interests or other interests that might be perceived to influence the results and/or discussion reported in this paper.




\begin{thebibliography}{10}


\bibitem[Adams and Laughlin(2001)]{2001Icar..150..151A} Adams, F.~C., Laughlin, G.\ 2001.\ Constraints on the Birth Aggregate of the Solar System.\ Icarus 150, 151–162. doi:10.1006/icar.2000.6567

\bibitem[Adams(2010)]{Adams2010} Adams, F.~C.\ 2010.\ The Birth Environment of the Solar System.\ Annual Review of Astronomy and Astrophysics 48, 47–85. doi:10.1146/annurev-astro-081309-130830

\bibitem[Arakawa and Kokubo(2023)]{2023A&A...670A.105A} Arakawa, S., Kokubo, E.\ 2023.\ Number of stars in the Sun's birth cluster revisited.\ Astronomy and Astrophysics 670. doi:10.1051/0004-6361/202244743



\bibitem[Batygin and Brown(2016)]{2016AJ....151...22B} Batygin, K., Brown, M.~E.\ 2016.\ Evidence for a Distant Giant Planet in the Solar System.\ The Astronomical Journal 151. doi:10.3847/0004-6256/151/2/22

\bibitem[Batygin et al.(2019)]{2019PhR...805....1B} Batygin, K., Adams, F.~C., Brown, M.~E., Becker, J.~C.\ 2019.\ The planet nine hypothesis.\ Physics Reports 805, 1–53. doi:10.1016/j.physrep.2019.01.009

\bibitem[Batygin et al.(2020)]{Bat20} Batygin, K., Adams, F.~C., Batygin, Y.~K., Petigura, E.~A.\ 2020.\ Dynamics of Planetary Systems within Star Clusters: Aspects of the Solar System's Early Evolution.\ The Astronomical Journal 159. doi:10.3847/1538-3881/ab665d

\bibitem[Batygin et al.(2021)]{Bat21} Batygin, K., Mardling, R.~A., Nesvorn{\'y}, D.\ 2021.\ The Stability Boundary of the Distant Scattered Disk.\ The Astrophysical Journal 920. doi:10.3847/1538-4357/ac19a4

\bibitem[Brasser et al.(2006)]{Brasser2006} Brasser, R., Duncan, M.~J., Levison, H.~F.\ 2006.\ Embedded star clusters and the formation of the Oort Cloud.\ Icarus 184, 59–82. doi:10.1016/j.icarus.2006.04.010

\bibitem[Brasser et al.(2012)]{Brasser2012} Brasser, R., Duncan, M.~J., Levison, H.~F., Schwamb, M.~E., Brown, M.~E.\ 2012.\ Reassessing the formation of the inner Oort cloud in an embedded star cluster.\ Icarus 217, 1–19. doi:10.1016/j.icarus.2011.10.012



\bibitem[Chambers(1999)]{1999MNRAS.304..793C} Chambers, J.~E.\ 1999.\ A hybrid symplectic integrator that permits close encounters between massive bodies.\ Monthly Notices of the Royal Astronomical Society 304, 793–799. doi:10.1046/j.1365-8711.1999.02379.x



\bibitem[Das and Batygin(2023)]{2023MNRAS.523.6103D} Das, A., Batygin, K.\ 2023.\ Suppression of the inclination instability in the trans-Neptunian Solar system.\ Monthly Notices of the Royal Astronomical Society 523, 6103–6113. doi:10.1093/mnras/stad1687



\bibitem[Fern{\'a}ndez(1997)]{Fernandez1997} Fern{\'a}ndez, J.~A.\ 1997.\ The Formation of the Oort Cloud and the Primitive Galactic Environment.\ Icarus 129, 106–119. doi:10.1006/icar.1997.5754


\bibitem[Grimm et al.(2022)]{2022ApJ...932..124G} Grimm, S.~L., Stadel, J.~G., Brasser, R., Meier, M.~M.~M., Mordasini, C.\ 2022.\ GENGA. II. GPU Planetary N-body Simulations with Non-Newtonian Forces and High Number of Particles.\ The Astrophysical Journal 932. doi:10.3847/1538-4357/ac6dd2



\bibitem[Hadden and Tremaine(2024)]{2024MNRAS.527.3054H} Hadden, S., Tremaine, S.\ 2024.\ Scattered disc dynamics: the mapping approach.\ Monthly Notices of the Royal Astronomical Society 527, 3054–3075. doi:10.1093/mnras/stad3478

\bibitem[Hills(1981)]{1981AJ.....86.1730H} Hills, J.~G.\ 1981.\ Comet showers and the steady-state infall of comets from the Oort cloud..\ The Astronomical Journal 86, 1730–1740. doi:10.1086/113058

\bibitem[Huang and Gladman(2024)]{2024ApJ...962L..33H} Huang, Y., Gladman, B.\ 2024.\ Primordial Orbital Alignment of Sednoids.\ The Astrophysical Journal 962. doi:10.3847/2041-8213/ad2686





\bibitem[Kaib and Quinn(2008)]{Kaib2008} Kaib, N.~A., Quinn, T.\ 2008.\ The formation of the Oort cloud in open cluster environments.\ Icarus 197, 221–238. doi:10.1016/j.icarus.2008.03.020



\bibitem[Madigan and McCourt(2016)]{2016MNRAS.457L..89M} Madigan, A.-M., McCourt, M.\ 2016.\ A new inclination instability reshapes Keplerian discs into cones: application to the outer Solar system.\ Monthly Notices of the Royal Astronomical Society 457, L89–L93. doi:10.1093/mnrasl/slv203


\bibitem[Miyamoto and Nagai(1975)]{1975PASJ...27..533M} Miyamoto, M., Nagai, R.\ 1975.\ Three-dimensional models for the distribution of mass in galaxies..\ Publications of the Astronomical Society of Japan 27, 533–543. https://ui.adsabs.harvard.edu/abs/1975PASJ...27..533M

\bibitem[Morbidelli(2002)]{2002mcma.book.....M} Morbidelli, A.\ 2002.\ Modern celestial mechanics : aspects of solar system dynamics.\ Modern celestial mechanics : aspects of solar system dynamics, by Alessandro Morbidelli. London: Taylor \& Francis, 2002, ISBN 0415279399.

\bibitem[Mu{\~n}oz-Guti{\'e}rrez et al.(2019)]{Munoz19} Mu{\~n}oz-Guti{\'e}rrez, M.~A., Peimbert, A., Pichardo, B., Lehner, M.~J., Wang, S.-Y.\ 2019.\ The Contribution of Dwarf Planets to the Origin of Jupiter Family Comets.\ The Astronomical Journal 158. doi:10.3847/1538-3881/ab4399




\bibitem[Nesvorn{\'y} et al.(2023)]{bib1} Nesvorn{\'y}, D., Bernardinelli, P., Vokrouhlick{\'y}, D., Batygin, K.\ 2023.\ Radial distribution of distant trans-Neptunian objects points to Sun's formation in a stellar cluster.\ Icarus 406. doi:10.1016/j.icarus.2023.115738

\bibitem[Nesvorn{\'y} et al.(2017)]{2017ApJ...845...27N} Nesvorn{\'y}, D., Vokrouhlick{\'y}, D., Dones, L., Levison, H.~F., Kaib, N., Morbidelli, A.\ 2017.\ Origin and Evolution of Short-period Comets.\ The Astrophysical Journal 845. doi:10.3847/1538-4357/aa7cf6






\bibitem[Saillenfest et al.(2019)]{2019A&A...629A..95S} Saillenfest, M., Fouchard, M., Ito, T., Higuchi, A.\ 2019.\ Chaos in the inert Oort cloud.\ Astronomy and Astrophysics 629. doi:10.1051/0004-6361/201936298

\bibitem[Sefilian and Touma(2019)]{2019AJ....157...59S} Sefilian, A.~A., Touma, J.~R.\ 2019.\ Shepherding in a Self-gravitating Disk of Trans-Neptunian Objects.\ The Astronomical Journal 157. doi:10.3847/1538-3881/aaf0fc


\bibitem[Tremaine(2023)]{2023MNRAS.522..937T} Tremaine, S.\ 2023.\ The Hamiltonian for von Zeipel-Lidov-Kozai oscillations.\ Monthly Notices of the Royal Astronomical Society 522, 937–947. doi:10.1093/mnras/stad1029





\bibitem[von Zeipel(1910)]{1910AN....183..345V} von Zeipel, H.\ 1910.\ Sur l'application des s{\'e}ries de M. Lindstedt {\`a} l'{\'e}tude du mouvement des com{\`e}tes p{\'e}riodiques.\ Astronomische Nachrichten 183, 345. doi:10.1002/asna.19091832202


\bibitem[Wisdom and Holman(1991)]{1991AJ....102.1528W} Wisdom, J., Holman, M.\ 1991.\ Symplectic maps for the N-body problem..\ The Astronomical Journal 102, 1528–1538. doi:10.1086/115978




\bibitem[Zderic and Madigan(2023)]{2023ApJ...948L...1Z} Zderic, A., Madigan, A.-M.\ 2023.\ Steeper Scattered Disks Buckle Faster.\ The Astrophysical Journal 948. doi:10.3847/2041-8213/accde2

\bibitem[Zink et al.(2020)]{2020AJ....160..232Z} Zink, J.~K., Batygin, K., Adams, F.~C.\ 2020.\ The Great Inequality and the Dynamical Disintegration of the Outer Solar System.\ The Astronomical Journal 160. doi:10.3847/1538-3881/abb8de

\end{thebibliography}



\end{document}